\documentclass{aa}
\usepackage{graphicx}
\def\deg{$^{\circ}$}

\global\def\antoe{Antonucci \cite{ant93}}

\global\def\baroo{Barth et al.\ \cite{bar99}}

\global\def\beaoop{Barth et al.\ (\cite{bar99})}
\global\def\braor{Braatz et al.\ \cite{bra94}}

\global\def\carie{Carter et al.\ \cite{car83}}

\global\def\claoi{Claussen et al.\ \cite{cla98}}
\global\def\claoip{Claussen et al.\ (\cite{cla98})}
\global\def\cohuq{Cohen et al.\ \cite{coh71}}

\global\def\daviy{Davies \& Illingworth \cite{dav86}}
\global\def\daviyp{Davies \& Illingworth (\cite{dav86})}
\global\def\diapq{Diamond et al.\ \cite{dia01}}

\global\def\fraoi{Frayer et al.\ \cite{fra98}}

\global\def\guaoo{Guainazzi \& Antonelli \cite{gua99}}
\global\def\guaoop{Guainazzi \& Antonelli (\cite{gua99})}
\global\def\guapp{Guainazzi et al.\ \cite{gua00}}
\global\def\guappp{Guainazzi et al.\ (\cite{gua00})}
\global\def\gabpp{Gabel et al.\ \cite{gab00}}
\global\def\gabppp{Gabel et al.\ (\cite{gab00})}

\global\def\hagppp{Hagiwara et al.\ (\cite{hag00})}
\global\def\heexx{Heeschen \cite{heeyi}, \cite{heeup}}

\global\def\heeie{Heeschen \& Puschell \cite{hee83}}

\global\def\jonir{Jones et al.\ \cite{jon84}}

\global\def\jeair{Jones et al.\ \cite{jon84}}

\global\def\jonpqp{Jones et al.\ (\cite{jon01})}

\global\def\kaspwp{Kaastra et al.\ (\cite{kas02})}

\global\def\keloi{Kellermann et al.\ \cite{kel98}}

\global\def\levot{Levinson et al.\ \cite{lev95}}

\global\def\knaoy{Knapp \& Rupen \cite{kna96}}
\global\def\knaoyp{Knapp \& Rupen (\cite{kna96})}
\global\def\maloy{Maloney et al.\ \cite{mal96}}

\global\def\meaoyp{Maloney et al.\ (\cite{mal96})}
\global\def\omapw{Omar et al.\ \cite{oma02}}
\global\def\omapwp{Omar et al.\ (\cite{oma02})}

\global\def\plaoi{Plana \& Boulesteix \cite{pla96}}

\global\def\pihppp{Pihlstr\"om et al.\ (\cite{pih00})}
\global\def\saruu{Sargent et al.\ \cite{sar77}}

\global\def\shoie{Shostak et al.\ \cite{sho83}}
\global\def\shoiep{Shostak et al.\ (\cite{sho83})}
\global\def\vaniy{van Gorkom et al.\ \cite{van86}}

\global\def\vanot{van Langevelde et al.\ \cite{van95}}
\global\def\vanotp{van Langevelde et al.\ (\cite{van95})}

\global\def\vanppp{van Langevelde et al.\ (\cite{van00})}

\global\def\walpp{Walker et al.\ \cite{wal00}}

\global\def\wanowp{Wang et al.\ (\cite{wan92})}
\global\def\weaoo{Weaver et al.\ \cite{wea99}}
\global\def\weaoop{Weaver et al.\ (\cite{wea99})}
\global\def\wikot{Wiklind et al.\ \cite{wik95}}
\global\def\wikotp{Wiklind et al.\ (\cite{wik95})}
\global\def\weaot{Wiklind et al.\ \cite{wik95}}
\global\def\weaotp{Wiklind et al.\ (\cite{wik95})}
\global\def\zenpw{Zensus et al.\ \cite{zen02}}

\newcommand\unit[1]{{\scriptsize [#1]}}
\begin{document}
\title{The Shroud Around the Twin Radio Jets in NGC\,1052}

\author{
   R.~C.~Vermeulen
   \inst{1}
   \and
   E.~Ros
   \inst{2}
   \and
   K.~I.~Kellermann
   \inst{3}
   \and
   M.~H.~Cohen
   \inst{4}
   \and
   J.~A.~Zensus
   \inst{2,3}
   \and
   H.~J.~van~Langevelde
   \inst{5}
}

\offprints{R.C.~Vermeulen,\hfill\break {\tt rvermeulen@astron.nl}}

\institute{
           Netherlands Foundation for Research in Astronomy, P.O. Box
           2, NL--7990 AA Dwingeloo, The Netherlands 
   \and
           Max-Planck-Institut f\"ur Radioastronomie,
           Auf dem H\"ugel 69, D-53121 Bonn, Germany 
   \and
           National Radio Astronomy Observatory, 520 Edgemont Road,
           Charlottesville, VA 22903, U.S.A. 
   \and
           California Institute of Technology, Pasadena, CA 91125, U.S.A. 
   \and
           Joint Institute for VLBI in Europe, P.O. Box 2, NL--7990
           AA Dwingeloo, The Netherlands 
}

\date{
Received 07 August 2002 / Accepted 25 November 2002 }

\abstract{

  We discuss multiple Very Long Baseline Interferometry (VLBI)
  continuum and spectral line imaging observations and Westerbork
  Synthesis Radio Telescope spectroscopy of the compact variable
  nuclear radio jet source in the elliptical galaxy \object{NGC\,1052}.
  Absorption and emission signatures reveal ionised, atomic, and
  molecular components of the surrounding medium.

  Ten epochs of Very Long Baseline Array (VLBA) data at 15\,GHz,
  spanning almost six years, show bi-symmetric jets, in which multiple
  sub-parsec scale features display outward motions of typically
  $v_{\rm app}\sim0.26c$ (${\rm H}_0=65$\,km\,s$^{-1}$\,Mpc$^{-1}$) on
  each side. The jets are most likely oriented near the plane of the
  sky.

  Multi-frequency VLBA observations at seven frequencies between 43 and
1.4\,GHz show free-free absorption in the inner parsec around the
nucleus, probably together with synchrotron self-absorption. The
free-free absorption is apparently due to a structure which is
geometrically thick and oriented roughly orthogonal to the jets, but
which is patchy. The western jet is covered more deeply and
extensively, and hence is receding.

  H{\sc i} spectral line VLBI observations reveal atomic gas in front
  of the approaching as well as the receding jet. There appear to be
  three velocity systems.  Broad, shallow absorption asymmetrically
  straddles the systemic velocity spanning $-35$ to
  85\,km\,s$^{-1}$. This gas could be local to the AGN environment, or
  distributed on galactic scales. Superimposed in the range 25 to
  95\,km\,s$^{-1}$ are several sharper (3--15\,km\,s$^{-1}$) features,
  each detectable over a few tenths of a pc at various places along the
  inner 2\,pc of the approaching jet. The third, deepest system is at
  ``high velocities", which is receding by 125 to 200\,km\,s$^{-1}$
  with respect to the systemic velocity of \object{NGC\,1052}.  It may
  have a continuous velocity gradient across the nucleus of some
  10\,km\,s$^{-1}$\,pc$^{-1}$. This atomic gas seems restricted to a
  shell 1--2\,pc away from the core, within which it might be largely
  ionised.

  Westerbork Synthesis Radio Telescope spectroscopy has revealed the
  18\,cm OH main lines (1667 and 1665\,MHz) in absorption along the
  full velocity span of $-35$ to 200\,km\,s$^{-1}$, with their line
  ratio varying roughly from 1:1 to 2:1. They are deepest in the high
  velocity system, where the OH profiles are similar to H{\sc i},
  suggesting co-location of that atomic and molecular gas, and leaving
  unclear the connection to the H$_2$O masing gas seen elsewhere.  In
  the high velocity system we have also detected the 18\,cm OH
  satellite lines: 1612\,MHz in absorption, and 1720\,MHz in
  emission. The conjugate behaviour of the satellite line profiles, and
  the variable main line ratio resemble the situation in \object{Cen~A}
  and \object{NGC\,253}.

\keywords{Galaxies: active -- Galaxies: jets -- Galaxies: nuclei --
  Galaxies: individual: NGC\,1052 -- Radio lines: galaxies}

}

\authorrunning{Vermeulen et al.\ }
\maketitle

\section{Introduction \label{sec:intro}}

The elliptical galaxy \object{NGC\,1052} is a low luminosity AGN which
has a well-studied LINER optical spectrum (e.g., \gabpp) and an
unusually prominent central radio source, with a flux density of
1--2\,Jy and a fairly flat spectrum between 1--30\,GHz, which is
variable on timescales of months to years (e.g., \heexx; \heeie). Since
\object{NGC\,1052} is comparatively nearby\footnote{We adopt
$H_0$=65\,km\,s$^{-1}$\,Mpc$^{-1}$, with no corrections for local
deviations from the Hubble flow, so that the optical stellar absorption
line heliocentric redshift, $cz=1474$\,km\,s$^{-1}$ or $z=0.0049$
(\saruu), corresponds to a distance of 22\,Mpc and implies that 1\,mas
corresponds to 0.1\,pc}, very detailed scrutiny of the active nucleus
and its inner environment are possible. This is important because the
extension to lower luminosity of the active nucleus/relativistic jet
phenomenon has not been well studied.

A 1.4\,GHz Very Large Array (VLA) image of \object{NGC\,1052} shows a
core-dominated radio structure, with only about 15\%\ of the flux
density in extended emission: there are two lobes spanning
$\sim3$\,kpc, possibly with hot spots (Wrobel \cite{wro84}).
The core luminosity at 5\,GHz is $1.3\times10^{23}$\,W\,Hz$^{-1}$
(\keloi, hereafter K98). Thus, while \object{NGC\,1052} is clearly in
the luminosity class of classical FR\,I radio sources, its radio
spectrum, morphology, and size (the entire radio source has dimensions
smaller than the host galaxy), suggest that it may be akin to
Gigahertz-Peaked-Spectrum or Compact-Steep-Spectrum sources, which are
generally thought to be young jets, propagating through, and perhaps
interacting with, a rich inner galactic medium (see the review by O'Dea
\cite{ode98}).

Early Very Long Baseline Interferometry (VLBI) observations showed that
most of the radio emission is from a sub-pc scale core (\cohuq),
extended over $\sim$15\,milliarcseconds (mas) in position angle
63\deg\ (\jonir). Recent work by Kadler et al.\ (\cite{kad02a})
shows that coincident X-ray, optical, and radio knots, at distances of
some 500\,pc on both sides of the core, are aligned with the
parsec-scale radio jets. The emission extended further out to a few kpc
seems to be more along a position angle of 95\deg, as already seen in
the VLA image by Wrobel (\cite{wro84}). \daviyp\ found the
parsec-scale jets to be approximately perpendicular to the gas
rotation axis, which they measured to be at $-41$\deg\ using the
optical emission lines. The same orientation is also seen in the
neutral, H{\sc i} 21\,cm line emitting gas component (\vaniy, hereafter
vG86). But, as is not uncommon in elliptical galaxies, the gas rotation
axis is oriented well away from that of the stars, which is at position
angle 25\deg, close to the projected photographic minor axis of the
galaxy (e.g, \daviy). \object{NGC\,1052} is probably a triaxial system,
and it is plausible that the gas, which has sub-solar abundances
(Sil'chenko \cite{sil95}) and a gas-to-dust ratio comparable to that in
our local Galaxy, was acquired as a result of one or more mergers or
tidal interactions, for example with the nearby spiral
\object{NGC\,1042} (see vG86; \daviy; \plaoi).

There is much observational evidence for the presence of a substantial
amount of gas and dust along the line of sight to the nucleus. First,
Chandra observations reveal an X-ray jet embedded into a 0.5\,keV
thermal plasma (Kadler et al.\ \cite{kad02a}).  Optical images show a
faint diffuse kpc-scale dust lane extending along the projected minor
axis (e.g., \carie; \daviy). Also, \gabppp\ have imaged an H$\alpha$
emission line filament on scales of $\sim$100\,pc. Towards the nucleus
there is reddening of the continuum light (e.g., \carie), as well as
the nebular emission lines (\gabpp). Polarised optical broad lines
(\baroo) give evidence for the presence along our direct line of sight
to the AGN of the kind of toroidal obscuring region often invoked in
unification models for various kinds of active galaxies (e.g.,
\antoe). The soft X-ray spectrum also shows strong absorption,
indicating a substantial, and possibly inhomogeneous column depth of
ionised gas towards the active nucleus (\guaoo; \guapp; \weaoo; Kadler
et al. \cite{kad02a}).

H{\sc i} 21\,cm line absorption has been found towards the bright
nuclear radio source at velocities ranging from
$\sim$1480--1680\,km\,s$^{-1}$, i.e.\ extending redward from systemic
over 200\,km\,s$^{-1}$ (\shoie; vG86). Around 1640\,km\,s$^{-1}$ H$_2$O
maser emission has been detected (\braor) and imaged with the VLBA
(\claoi). Recently, 1667 and 1665\,MHz OH absorption near
1640\,km\,s$^{-1}$ has also been detected (\omapw).  The presence of CO
lines is unclear: \wanowp\ claimed a detection of CO(1--0) in emission
near 1420\,km\,s$^{-1}$, but the peak strength of 150\,mK is much above
the 20 mK 3$\sigma$ upper limit claimed by \wikotp, and is also not
confirmed by \knaoyp. Meanwhile, \knaoyp\ did find tentative absorption
features in CO(1--0) in the velocity range 1550--1700\,km\,s$^{-1}$.
As they point out, lines at that velocity might also be visible in the
CO(1--0) and CO(2--1) spectra of \weaotp. Since there is now confirmed
molecular gas at this velocity in OH and also in H$_2$O masers, it is
important to verify the CO results.

This paper deals with various aspects of the radio jets and the
measurements of ionised atomic and molecular gas in their surrounding
medium.  Section~\ref{sec:obs} collects the details of all of the
observations and the data reduction. In Sect.~\ref{sec:kinematics} we
discuss the morphology and kinematics of sub-pc scale features in the
jets, based on 15\,GHz VLBA observations made at a total of ten
different epochs, spanning a period of more than five years. In
Sect.~\ref{sec:iongas} we present the broad-band continuum spectra of
various jet components, which indicate extensive, asymmetric free-free
absorption; this is based on VLBA observations at seven frequencies,
spanning the range from 1.4\,GHz up to 43\,GHz. Kellermann et al.\
(\cite{kel99}, hereafter K99) reported on motions and free-free
absorption in \object{NGC\,1052}, now fully described in this
paper. Kameno et al.\ (\cite{kam01}, hereafter Ka01) later
analysed another dataset also showing free-free absorption.  Kadler et
al.\ (\cite{kad02a}; \cite{kad02b}) analysed further multi-frequency
data finding free-free absorption also, and linearly polarised emission
at 5\,GHz.  Section~\ref{sec:atogas} shows the spatial and velocity
distribution of the H{\sc i} absorption on sub-pc scales, based on
H{\sc i} 21\,cm spectral line VLBI
observations. Section~\ref{sec:molgas} deals with Westerbork Synthesis
Radio Telescope (WSRT) observations of all four 18\,cm OH lines, using
the new wide band multi-channel digital backend (DZB) which is very
well suited for observing spectral lines over wide bandwidths with
multiple spectral channels.  We find the OH 1665 and 1667\,MHz lines
over a much wider velocity range than originally discovered by \omapwp,
and we also find for the first time the presence of the OH 1612 and
1720\,MHz lines, in absorption and emission, respectively.

\section{The observations and data reduction \label{sec:obs}}

Table~\ref{table:obstab} gives an overview of all observations of
\object{NGC\,1052} which we discuss in this paper. These include
VLBA multi-epoch 15\,GHz continuum monitoring, VLBA multi-frequency
continuum imaging, VLBI H{\sc i} 21\,cm line imaging, and WSRT OH
18\,cm line spectroscopy.

\subsection{Multi-epoch VLBA monitoring \label{subsec:obs-multiep}}

We first recognised \object{NGC\,1052} as a source of special interest
from our 15\,GHz (2\,cm) VLBA study of the morphology and kinematics of
strong compact radio sources (K98; \zenpw) when we saw the
striking and unusual bi-symmetry of the parsec-scale jets. Thus, since
1995, we have included \object{NGC\,1052} in most of our multi-snapshot
15\,GHz VLBA survey observing sessions. Sources are typically observed
for 4--8 minutes per hour during 8-hour blocks, using all antennas of
the VLBA where the source is above 10$^{\circ}$ elevation. The survey
observations and data reduction are further described in K98, and that
paper contains the 15\,GHz image of \object{NGC\,1052} obtained at
epoch 1997.20. We have also obtained some deeper 15\,GHz VLBA images
from dedicated observing runs on \object{NGC\,1052}.  All ten 15\,GHz
images are displayed in Fig.~\ref{fig:tenims}. The kinematics are
discussed in Sect.~\ref{sec:kinematics}.

\begin{table}
    \caption{Overview of all datasets discussed in this paper.}
\begin{center}
\begin{footnotesize}
\begin{tabular}{@{}l@{~}c@{~}c@{~}p{16mm}@{~}p{33mm}@{}}
\hline
\noalign{\smallskip}
\hline
\noalign{\smallskip}
Date       & $t_{\rm int}$  & $\nu$ & Array  &     Remarks \\
\unit{yyyy/mm/dd}
           & \unit{min}  & \unit{GHz} &        &             \\
\noalign{\smallskip}
\hline
\noalign{\smallskip}
1995/07/28 &   40 & 15.4 & {\footnotesize VLBA   }&     15\,GHz monitoring     \\
1995/12/16 &   40 & 15.4 & {\footnotesize VLBA   }&     15\,GHz monitoring     \\
1996/07/10 &   24 & 15.4 & {\footnotesize VLBA   }&     15\,GHz monitoring    \\
1996/10/24 &  120 & 15.4 & {\footnotesize VLBA   }&     Continuum imaging     \\
1997/03/13 &  100 & 15.4 & {\footnotesize VLBA   }&     15\,GHz monitoring     \\
1997/07/09 &   62 &  5.0 & {\footnotesize VLBA   }&     Continuum imaging     \\
1997/07/09 &   62 &  8.4 & {\footnotesize VLBA   }&     Continuum imaging     \\
1997/07/09 &  100 & 15.4 & {\footnotesize VLBA   }&     Continuum imaging     \\
1997/07/12 &  120 & 22.2 & {\footnotesize VLBA+Y1}&     Continuum imaging     \\
1997/07/12 &  180 & 43.2 & {\footnotesize VLBA+Y1}&     Continuum imaging     \\
1997/07/15 &  410 &  1.4 & {\footnotesize VLBA   }&     Contin.\ + line imaging     \\
1997/07/15 &  410 &  1.7 & {\footnotesize VLBA   }&     Contin.\ + line imaging     \\
1998/02/01 &  100 & 15.4 & {\footnotesize VLBA   }&     Continuum imaging     \\
1998/07/19 &  570 &  1.4 & {\footnotesize VLBA+ Y27+Eb}& Line imaging     \\
1998/11/01 &   36 & 15.4 & {\footnotesize VLBA   }&     15\,GHz monitoring     \\
1999/11/07 &   44 & 15.4 & {\footnotesize VLBA   }&     15\,GHz monitoring     \\
2001/03/15 &   45 & 15.4 & {\footnotesize VLBA   }&     15\,GHz monitoring     \\
2001/12/22 &  585 &  1.7 & {\footnotesize WSRT   }&     Spectrum OH-main     \\
2002/01/07 &  240 &  1.7 & {\footnotesize WSRT   }&     Spectrum OH-main     \\
2002/02/16 &  420 &  1.7 & {\footnotesize WSRT   }&     Spectrum 1720     \\
2002/02/18 &  210 &  1.7 & {\footnotesize WSRT   }&     Spectrum 1720     \\
2002/02/19 &  380 &  1.7 & {\footnotesize WSRT   }&     Spectrum 1720     \\
2002/02/23 &  250 &  1.6 & {\footnotesize WSRT   }&     Spectrum 1612     \\
2002/02/24 &  510 &  1.6 & {\footnotesize WSRT   }&     Spectrum 1612     \\
\noalign{\smallskip}
\hline
\end{tabular}
\end{footnotesize}
\end{center}
\label{table:obstab}
\end{table}

\subsection{Multi-frequency VLBA imaging \label{subsec:obs-multifreq}}

We report on nearly simultaneous multi-frequency VLBA continuum
observations of \object{NGC\,1052}, made on 1997 July 9 at 5, 8, and
15\,GHz, and on 1997 July 12 at 22 and 43\,GHz; for the latter, a
single VLA antenna (``Y1") was added to the VLBA in order to improve
short baseline coverage. The observing frequency was changed every
5--10 minutes.  Both left- and right-circular polarisation were
recorded with 2-bit samples over 4 contiguous 8\,MHz bands; the
record/no-record duty cycle was 50\%, since there was a time-averaged
total data rate limit of 128\,Mbit\,s$^{-1}$. On 1997 July 15, a 4\,MHz
wide band at 1411.204--1415.204\,MHz, targeting the H{\sc i} 21\,cm
line, was recorded continuously with 2-bit samples at 16
Megasamples\,s$^{-1}$ (twice the Nyquist rate), in both hands of
circular polarisation. This setup optimises the sensitivity for
limited-bandwidth signals. We have obtained excellent H{\sc i} spectral
line data, as well as a 1.4\,GHz continuum image from the line-free
channels, as described below. Simultaneously, using separate local
oscillator (LO) settings in other base-band converter (BBC) channels,
we also alternated every 5--10 minutes between recording
1654.958--1658.958\,MHz and 1656.945--1660.945\,MHz, in the same mode
as for the 1.4\,GHz channels, in order to search for possible OH
lines. We have since discovered that this setup would not yield
adequate sensitivity for OH spectral line imaging (see
Sect.~\ref{sec:molgas}), and we have simply combined all of the data to
yield a 1.6\,GHz continuum image. All continuum datasets were reduced
in standard fashion (e.g., K98): amplitude calibration and
fringe-fitting in {\sc aips}, followed by editing, iterative
self-calibration and cleaning in {\sc difmap}.  Special care was taken
during fringe-fitting at 43\,GHz to obtain as many secure detections on
long baselines as possible. The continuum images are displayed in
Fig.~\ref{fig:ffims}, and the resultant broad-band continuum spectral
information is analysed in Sect.~\ref{sec:iongas}.  We identified
"components" in the images which appeared at the same place at all (or
most) wavelengths, and fitted Gaussians at these fixed locations to the
visibility data.  The component locations are shown on top of the
images in Fig.~\ref{ffmods}.  Due to spectral effects as well as
differences in resolution, it was not possible to find all components
at all epochs, but it was always possible to obtain a unique relative
registration through the cross-identification of several components in
both jets between adjacent frequencies, working gradually from the
centre outward at progressively lower frequencies. The determination of
relative flux densities of features observed at the different
frequencies of course suffers from all of the usual VLBI imaging
problems, ranging from inaccurate, time variable {\it a priori} gain
factors and system temperatures, through floating gains incurred during
self-calibration, to flux density lost from (over)resolved structures.
Conservatively, we would caution against over-interpretation of
phenomena which would require relative flux density measurements
accurate to better than 10--20\%.  Further details on the
multi-frequency analysis and the kinematics are given in
Figs.~\ref{fig:ffcomp}, \ref{fig:ffflux}, and \ref{fig:ffrad}.

\subsection{VLBI H{\sc i} line observations \label{subsec:obs-hi}}

The 1997 July 15 data spanned 1411.204--1415.204\,MHz, which
covers about 850\,km\,s$^{-1}$, centred on $v_{\rm
opt,hel}=1550$\,km\,s$^{-1}$ for the H{\sc i} 21\,cm line. The data
were correlated with 4 sec accumulation intervals and 512 spectral
channels over the 4\,MHz wide band, yielding a spectral resolution of
2\,km\,s$^{-1}$. On 1998 July 19, we obtained a full-track 1.4\,GHz
VLBI spectral line observation, using the ten antennas of the VLBA,
together with the VLA operating as a tied array (``Y27"), and the
Effelsberg radio telescope (``Eb"). Unfortunately, we found that for
Effelsberg and the Mauna Kea VLBA antenna, which are rather isolated on
opposite ends of the VLBI array, uncertainties in the self-calibration
were too large to make a reliable contribution to the final results. In
this second observation we simultaneously used 8\,MHz and 1\,MHz wide
bands, 1409.01--1417.01\,MHz and 1412.51--1413.51\,MHz, co-centred on
$v_{\rm opt,hel}=1595$\,km\,s$^{-1}$, in the two senses of circular
polarisation, recorded with 2-bit samples at 16\,Megasamples\,s$^{-1}$
each (a factor 8 oversampled w.r.t.\ the Nyquist rate for the narrow
bandwidth). Correlation using 16 second accumulation intervals took
place with 1024 spectral channels, yielding spectral resolutions of
2\,km\,s$^{-1}$ (the same as for the 1997 dataset) and
0.25\,km\,s$^{-1}$, respectively. Initial reduction of the spectral
line datasets took place in {\sc aips}: after {\it a priori} amplitude
calibration and fringe fitting, complex passband solutions for each of
the antennas were obtained from the strong sources \object{3C\,454.3}
and \object{3C\,84}, observed immediately preceeding and following
\object{NGC\,1052}. After application of the solutions, the two
parallel hands of polarisation were added (cross-polarisation
correlation did not take place). All further analysis took place in
{\sc difmap}.  The spectral channels well away from the absorption line
complex were averaged and used for continuum self-calibration and
imaging. The dataset with 1\,MHz bandwidth from 1998 has no line-free
channels, and so the continuum self-calibration solutions from the
co-centred 8\,MHz wide band were adopted.  The resultant 1997 1.4\,GHz
image is shown in Fig.~\ref{fig:ffims}, and this dataset was used in
the multi-frequency analysis presented in Sect.~\ref{sec:iongas}. The
antenna phase corrections as a function of time derived from the
continuum were then applied to all line channels. Spectra at different
positions along the jets were determined from Gaussian components
fitted directly to the visibility data, guided by the desire for an
optimum discrimination between the locations of the various absorption
line components. The model was first developed using the integrated
continuum channels, and the flux densities of the model components were
then re-fitted separately for each spectral line channel, while their
positions and sizes were kept fixed. We have determined that the
results are stable against reasonable perturbations, for example,
slightly altered model positions, or using different sliding averages
over the line channels. The resultant spectra are shown in
Fig.~\ref{hispec97} and Fig.~\ref{hispec98} for those components where
the spectral signal-to-noise ratio was useful.

\subsection{WSRT OH line observations \label{subsec:obs-oh}}

Triggered by the discovery of the 18\,cm OH main lines (1665
and 1667\,MHz) near $v=1640$\,km\,s$^{-1}$ by \omapwp, we have used the
new, wide-band, multi-channel backend (DZB) at the WSRT to obtain high
quality spectra to probe the presence of all four 18\,cm OH lines
(1612, 1665, 1667, and 1720\,MHz) over the full velocity range seen in
H{\sc i}.  A number of scans were obtained in the period December 2001
-- February 2002 (see table~\ref{table:obstab}).  The main lines are
sufficiently close together in frequency to observe jointly; we have
used a 10\,MHz wide band which probes a wide velocity range, as
discussed in Sect.~\ref{sec:molgas}.  Each of the satellite lines (1612
and 1720\,MHz) was observed separately with a 5\,MHz wide band. In all
cases there were 512 spectral channels over the band (half the nominal
DZB capacity), Hamming tapered to avoid spectral ringing of radio
frequency interference (RFI), should it occur.  The resultant spectral
resolution is near 7\,km\,s$^{-1}$ for the main lines, and around
3.5\,km\,s$^{-1}$ for the satellite lines. We used 2-bit sampling, and
a data accumulation interval of 60 sec. Initial phase, flux density,
and passband calibration was based on the strong, unresolved sources
\object{3C\,48} and \object{3C\,147}, which were always observed
immediately before and after \object{NGC\,1052}. The spectral slope
remaining in Fig.~\ref{ohmain} is attributable to unmodeled spectral
index differences between the calibrators and \object{NGC\,1052}, and
the flux density scales are accurate to 5\%\ or better. We added the
two orthogonal linear polarisations (no cross-polarisation correlation
was carried out) after separate calibration. Self-calibration on the
average of the line-free channels was then performed to find
time-dependent complex gain corrections, and these were applied to the
visibilities in each of the spectral channels.  \object{NGC\,1052} is
by far the dominant point source, in the centre of the field, and the
final spectra were obtained simply as a vector average over all times
and interferometers. The angular resolution of the WSRT is around 13
arcsec, and we did verify that the results are not affected by the
presence of the lobes, containing less than 50\,mJy each, and situated
at 11 and 14 arcsec to the east and west, respectively. The data for
the line with rest frequency 1612\,MHz were significantly affected by
RFI emitted near 1603 and 1605\,MHz, possibly due to GLONASS
satellites. Significant stretches of time had to be edited out, and all
data on baselines shorter than 750\,m were completely removed. Even so,
for this line the noise at velocities below 1500\,km\,s$^{-1}$ is
dominated by residual RFI.

\section{Kinematics \label{sec:kinematics}}

\begin{figure}[h!]
    \begin{center}
    \includegraphics[width=8.3cm,angle=0]{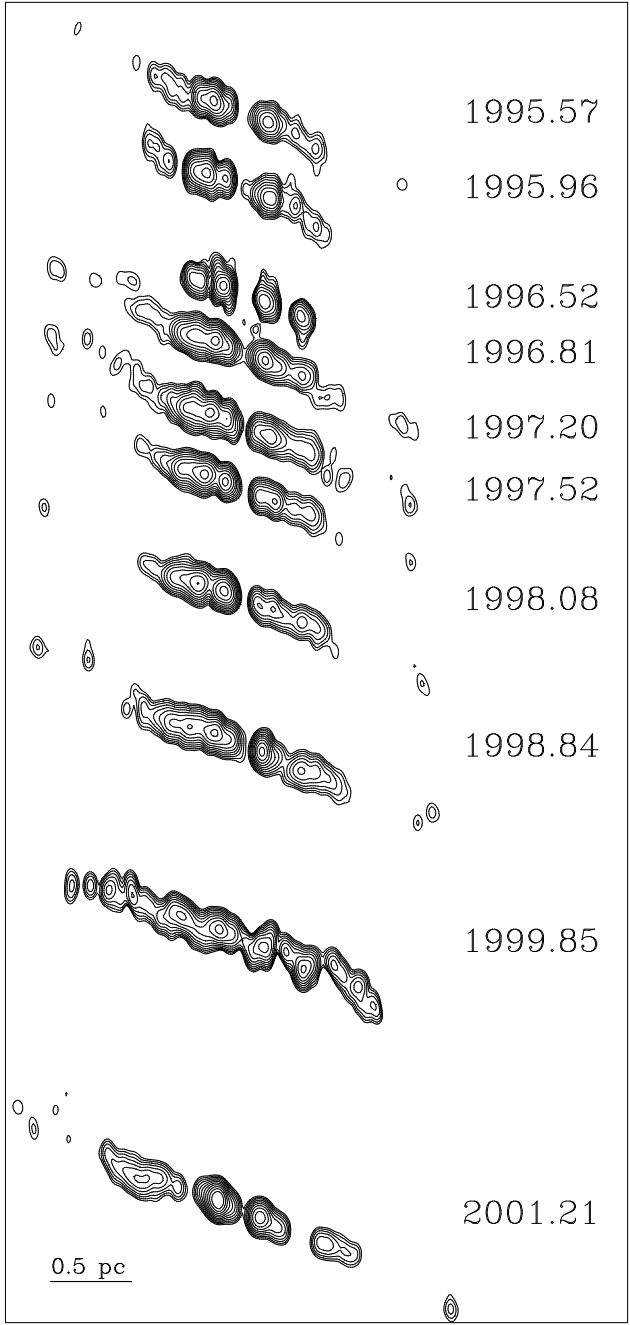}
    \caption{
      VLBA 15\,GHz contour images of \object{NGC\,1052} at ten epochs,
      aligned following the kinematic analysis of
      Sect.~\ref{sec:kinematics}, and shown spaced by their relative
      time intervals (details in Table~\ref{table:obstab}). The contour
      levels all increase by factors of $3^{1/2}$, starting at 0.58\%
      from the peak of brightness in each epoch, except in 2001.21,
      where it starts at 0.19\%.  The common convolving restoring beam
      is 0.10$\times$0.05\,pc in position angle 0\degr.
            }
\label{fig:tenims}
\end{center}
\end{figure}

\begin{figure}[htbp]
    \begin{center}
    \includegraphics[width=8.3cm,angle=0]{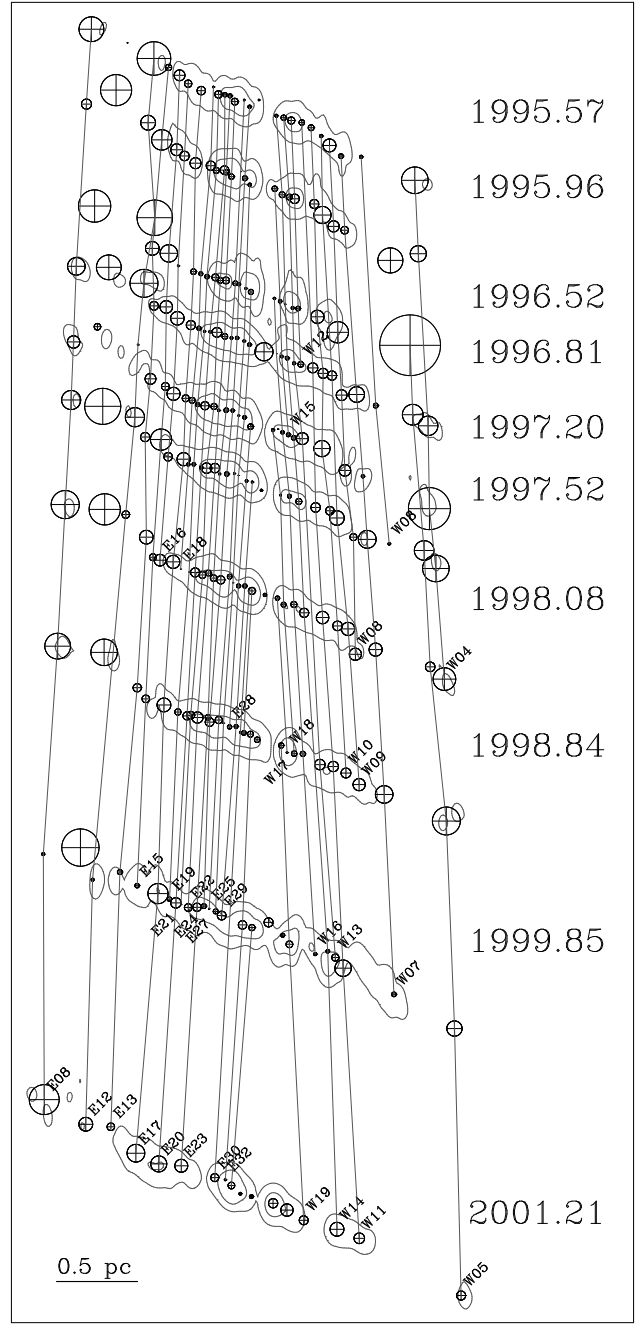}
    \caption{
      Illustration of the locations, sizes, and motions of the
      Gaussian model components used for the kinematic analysis in
      Sect.~\ref{sec:kinematics}. The images are identical to those
      shown in Fig.~\ref{fig:tenims}, but with only the contour levels
      of 0.58\% and 15.6\% (0.19 and 5.2\% in 2001.21) shown.  The
      components are labelled with increasing number inwards in the
      eastern (E) and western (W) jets (see text).  }
\label{fig:modims}
\end{center}
\end{figure}

Ten epochs of VLBA data at 15\,GHz have been accumulated, spanning the
period 1995 July - 2001 March. The relative alignment of the images,
shown in Fig.~\ref{fig:tenims}, is explained below. At all epochs the
source appears symmetric with oppositely directed, slightly curved
jets. Note the prominent gap 0.1--0.2\,pc west of the brightest feature
in most images. From the equivalent colour images a movie has been
produced, which can be viewed at the 2-cm Survey web site, {\tt
http://www.nrao.edu/2cmsurvey}.  The frames in this movie were
generated by linear interpolation of the {\sc clean} models over time
at every pixel between the ten epochs of observation. Note that this
procedure will create a non-physical appearance whenever features move
by more than half a synthesised beam width between epochs; this is
indeed the case several times during the sequence.

The ten 15\,GHz images clearly reveal motions away from the centre of
symmetry. However, at the high linear resolution we have achieved, this
source shows complex structure and evolution, not easily described by a
simple collection of discrete components moving with well-defined
velocities throughout the nearly six-year time period spanned by our
observations. The flux density of features evolves rapidly, on time
scales of months, and new features seem to arise frequently in both
inner jets. There is thus the risk of stroboscopic effects whenever the
time coverage is insufficiently dense; indeed, we have omitted 2001.21,
the last 15\,GHz epoch, from our motion analysis because we are not
sure we can unambiguously tie in the evolution of the jet emission over
the sixteen-month interval to the previous epoch.  A quantitative
analysis of the jet flow is also made more difficult by the differing
$(u,v)$ coverage and varying sensitivity among epochs. These are due to
the effects of weather and to occasional technical failures at one or
more telescopes.

\begin{figure*}[htbp]
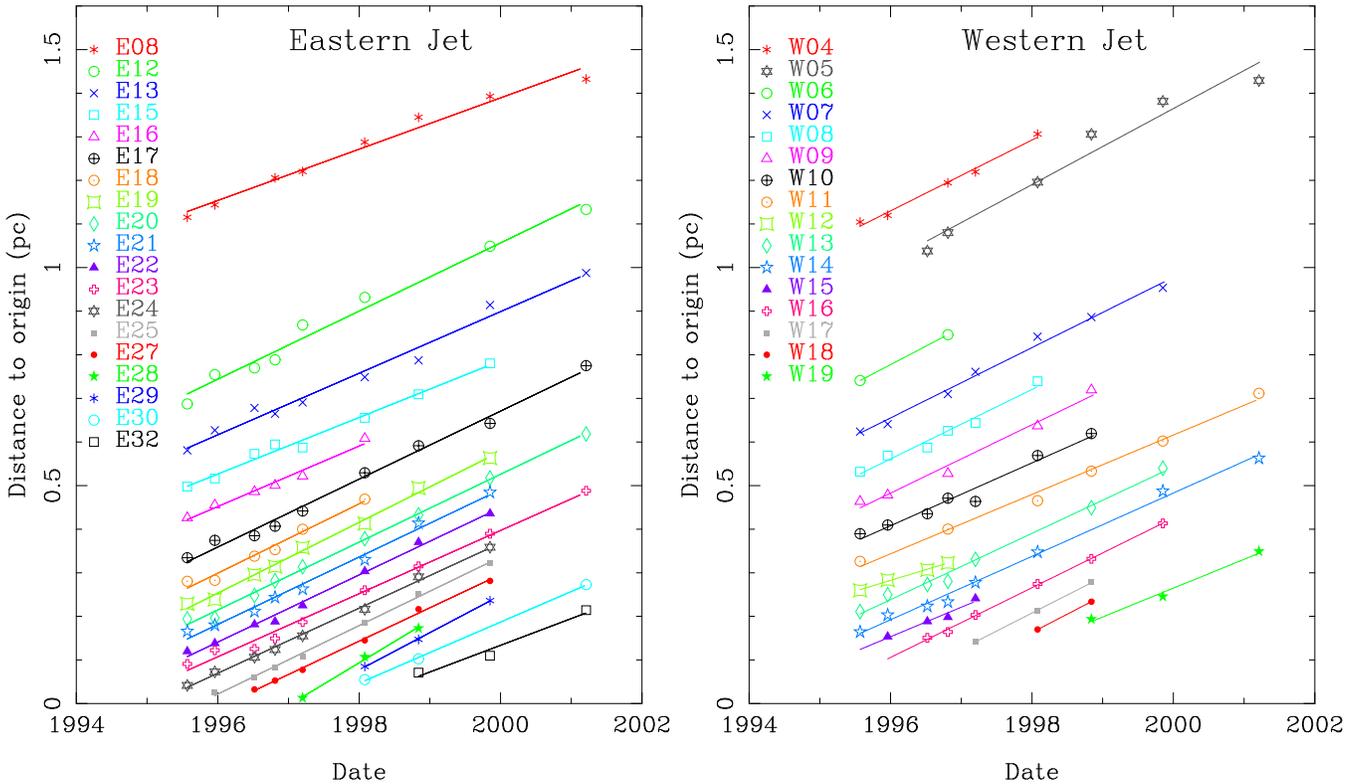

    \begin{center}
    \resizebox{\hsize}{!}{
    \includegraphics[width=9cm,angle=0]{H3907F3A.ps}
    \includegraphics[width=9cm,angle=0]{H3907F3B.ps}
}
    \caption{
        Distances to the adopted origin (see
        Sect.~\ref{sec:kinematics}) for the Gaussian components fitted
        to the 15\,GHz datasets at ten epochs (see
        table~\ref{table:veltab}).  Lines show the best-fit linear
        velocities.
            }
\label{fig:modrad}
\end{center}
\end{figure*}

\begin{figure*}[htbp]
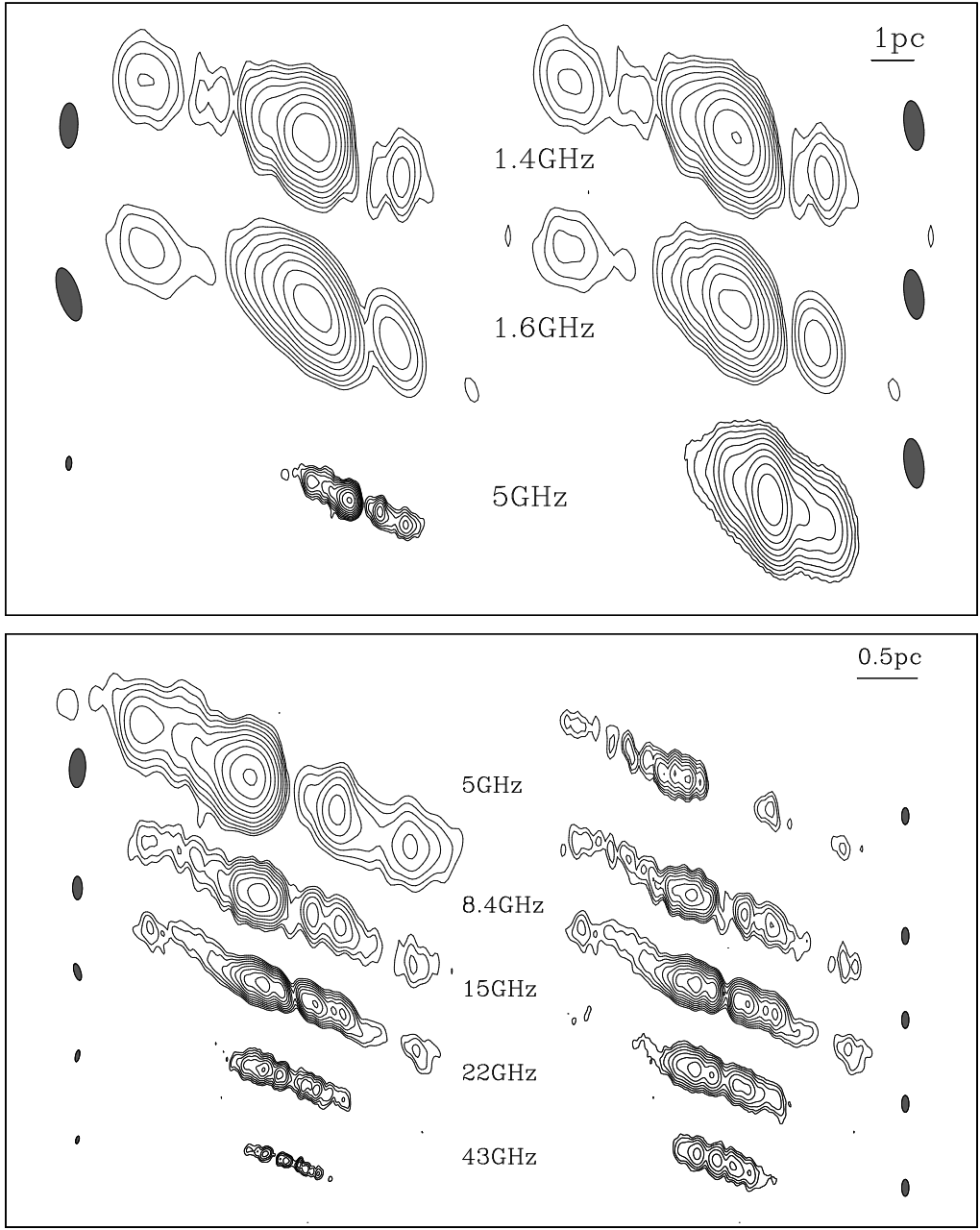

    \begin{center}
    \includegraphics[width=13.5cm,angle=0]{H3907F4A.ps}

    \vspace{5pt}

    \includegraphics[width=13.5cm,angle=0]{H3907F4B.ps}

    \caption{
        Contour images of \object{NGC\,1052} for the 1997.52 epoch at
        43, 22, 15, 8, 5, 1.6, and 1.4\,GHz. The left-hand side shows
        each image at a resolution appropriate to its wavelength, and,
        to facilitate inter-comparison, in the right-hand column the
        restoring beam is 1.18$\times$0.46\,pc for the lower
        frequencies (top), and 0.14$\times$0.06\,pc for the higher
        frequencies (bottom); the 5\,GHz image is repeated in both
        panels. The lowest contour levels are 0.83, 1.21, 1.57, 3.44,
        1.04, 3.05, and 7.44\,mJy\,beam$^{-1}$, and in all cases they
        increase by a factor of $2^{1/2}$.
            }
\label{fig:ffims}
\end{center}
\end{figure*}

Nevertheless, we have achieved a reliable quantitative motion analysis,
which helps to constrain the geometry of the jets and surrounding
medium in \object{NGC\,1052}. For each epoch we used {\sc difmap} to
carry out an iterative least-squares fit to the visibility data of a
series of circular Gaussian features which reasonably approximate the
actual images, as verified by inspection of the residuals between the
visibility data and the model in the image plane. The development of a
component set consistent over time was guided by extensive visual
cross-comparisons of the ten images (Fig.~\ref{fig:tenims}). Many of
the components found at 15\,GHz are also seen in the higher resolution
43\,GHz image which is shown in Fig.~\ref{fig:ffims}. While not every
component should be interpreted as a true physical entity (a plasmon or
a ``cannon-ball''), we are confident that the set as a whole is an
appropriate representation of many morphological features which can be
distinctly tracked as they persist for multiple epochs. These features
are labelled in Figs.~\ref{fig:modims} and \ref{fig:modrad}.

In order to solve for the motions of the components with respect to the
origin of ejection, one should ideally know beforehand the location of
the origin at each epoch. However, our observations were not
phase-referenced to a celestial position calibrator
source. Furthermore, as is shown in Sect.~\ref{sec:iongas}, the origin
of ejection is likely to be in the central emission gap, and therefore
there is no recognisable, persistent, stationary emission feature on
which to align the models and images.  But, as demonstrated by
Figs.~\ref{fig:modims} and \ref{fig:modrad}, we have found that a
plausible {\it relative} alignment of the epochs exists, in which the
motions of most of the components, on both sides of the centre, are
consistent with being {\it linear} on the sky and {\it constant} over
time. The jets are clearly curved, and this is reflected by the fact
that we have found components moving along different position
angles. While a simple ballistic model fits well, we cannot rule out
that mild accelerations and decelerations or bending may occur as the
components traverse a significant fraction of a parsec over a number of
years.  In order to determine that, a longer monitoring is required.

\begin{table}
    \caption{
        Overview of the velocities of all components analysed in
        Sect.~\ref{sec:kinematics}.
            }
\begin{center}
\begin{tabular}{@{}lccc@{}}
\hline
\noalign{\smallskip}
\hline
\noalign{\smallskip}
Comp.\
    &    $v_x$  &    $v_y$ & $(v_x^2+v_y^2)^{1/2}$ \\
ID  &  \unit{pc\,yr$^{-1}$}
    &  \unit{pc\,yr$^{-1}$}
    &  \unit{pc\,yr$^{-1}$} \\
\noalign{\smallskip}
\hline
\noalign{\smallskip}
E08 &   0.067  &  0.013   &  0.068  \\
E12 &   0.090  &  0.006   &  0.090  \\
E13 &   0.062  &  0.032   &  0.070  \\
E15 &   0.063  &  0.020   &  0.066  \\
E16 &   0.067  &  0.017   &  0.069  \\
E17 &   0.066  &  0.036   &  0.075  \\
E18 &   0.084  &  0.011   &  0.084  \\
E19 &   0.074  &  0.033   &  0.081  \\
E20 &   0.074  &  0.027   &  0.079  \\
E21 &   0.073  &  0.029   &  0.079  \\
E22 &   0.070  &  0.033   &  0.078  \\
E23 &   0.071  &  0.029   &  0.076  \\
E24 &   0.070  &  0.036   &  0.078  \\
E25 &   0.068  &  0.045   &  0.081  \\
E27 &   0.069  &  0.043   &  0.082  \\
E28 &   0.080  &  0.058   &  0.099  \\
E29 &   0.081  &  0.045   &  0.093  \\
E30 &   0.071  &  0.038   &  0.080  \\
E32 &   0.049  &  0.028   &  0.056  \\
\noalign{\smallskip}
\hline
\noalign{\smallskip}
W19 &  $-$0.050  & $-$0.022 &    0.055  \\
W18 &  $-$0.079  & $-$0.061 &    0.100  \\
W17 &  $-$0.074  & $-$0.023 &    0.078  \\
W16 &  $-$0.077  & $-$0.022 &    0.080  \\
W15 &  $-$0.065  & $-$0.029 &    0.072  \\
W14 &  $-$0.074  & $-$0.013 &    0.075  \\
W13 &  $-$0.074  & $-$0.018 &    0.076  \\
W12 &  $-$0.046  & $-$0.020 &    0.050  \\
W11 &  $-$0.061  & $-$0.025 &    0.066  \\
W10 &  $-$0.067  & $-$0.026 &    0.072  \\
W09 &  $-$0.072  & $-$0.026 &    0.077  \\
W08 &  $-$0.067  & $-$0.052 &    0.085  \\
W07 &  $-$0.080  & $-$0.018 &    0.082  \\
W06 &  $-$0.087  & $-$0.033 &    0.094  \\
W05 &  $-$0.087  & $-$0.065 &    0.109  \\
W04 &  $-$0.071  & $-$0.041 &    0.082 \\
\noalign{\smallskip}
\hline
\end{tabular}
\end{center}
\label{table:veltab}
\end{table}

For ballistic motions, one can treat the orthogonal $(x,y)$ velocities
on the plane of the sky separately. Using all of the component
positions at all of the epochs (except 2001.21) together, we have
therefore performed a single least-squares fit to solve for relative
$(x,y)$ offsets on the sky between all epochs, simultaneously with
linear $(x,y)$ velocities for all individual
components. Table~\ref{table:veltab} gives the resultant velocities for
each of the components identified in Fig.~\ref{fig:modims}. In
Fig.~\ref{fig:modrad} we show the measured component positions, with
the best-fit relative inter-epoch alignments applied. The lines show
the best-fit linear velocity of each component for which we had
confidence in the cross-epoch identification. At most epochs we
detected additional radio emitting features. While appearing to move
outward with roughly similar velocities, these features are too diffuse
or too faint to have their position tightly constrained, and we have
omitted them from the simultaneous least-squares fit.

The absolute location (0,0) of the stationary kinematic origin, or
core, is in principle given by the intersection of the
back-extrapolated ballistic trajectories, since the jets are somewhat
curved. However, we have found that this procedure is rather sensitive
to the chance configuration of available components. We estimate that
the absolute location of the centre used in the figures and tables is
uncertain by about $\pm0.2$\,mas (0.02\,pc), although the relative
alignment between the epochs is much better. We find that our
multi-epoch registration (not available for the data studied by Ka01)
provides an indispensable basis for an accurate study of the detailed
geometry of the radio source and the enshrouding absorption
(Sect.~\ref{sec:iongas},~\ref{sec:atogas}), with respect to a
well-defined kinematic centre.

\begin{figure}[htbp]
    \begin{center}
    \includegraphics[width=8.5cm,angle=0]{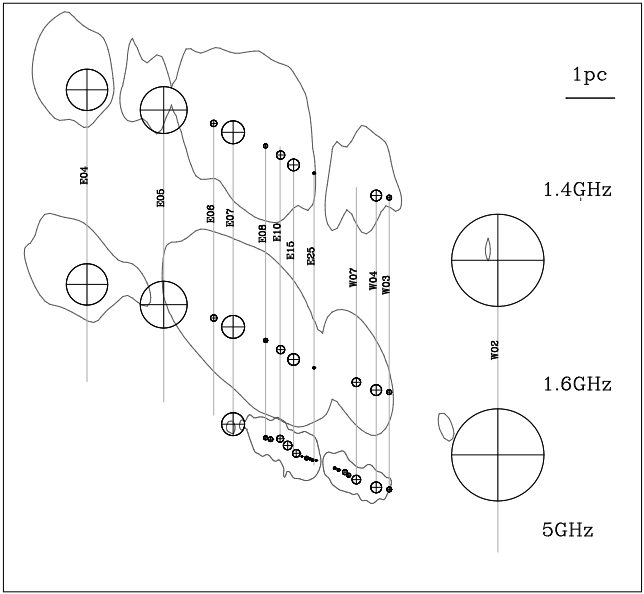}

\vspace{5pt}

    \includegraphics[width=8.5cm,angle=0]{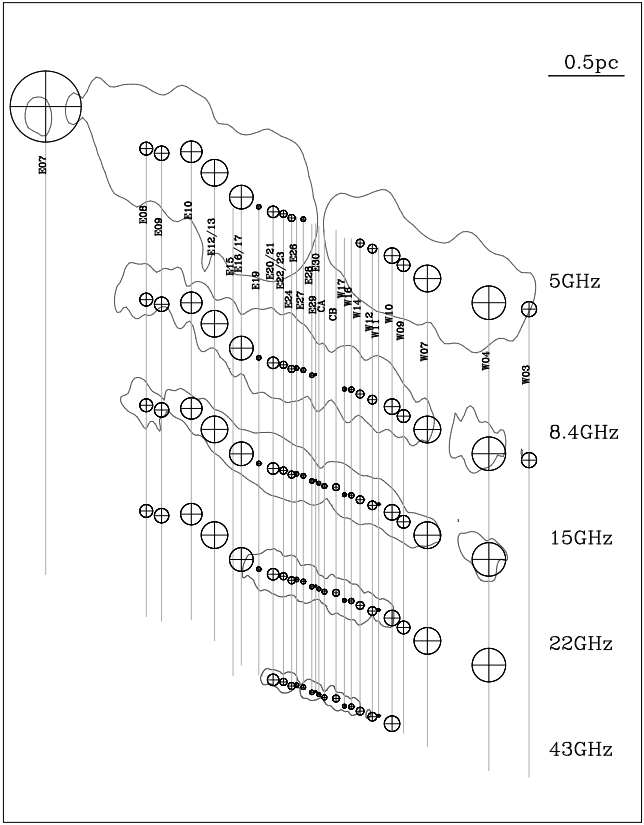}
    \caption{
      Illustration of the locations and sizes of the Gaussian
      model components used for the multi-frequency analysis in
      Sect.~\ref{sec:iongas}. The images are identical to those shown
      in Fig.~\ref{fig:ffims}, but with only the lowest contour level
      from those images shown.  }
\label{ffmods}
\end{center}
\end{figure}

The data establish quite clearly that the features on the two sides
move in opposite directions with roughly equal apparent velocities of
$0.78\pm0.12$\,mas\,yr$^{-1}$ ($0.078\pm0.012$\,pc\,yr$^{-1}$), which
corresponds to a linear velocity of 0.26$\pm$0.04$c$. These apparent
velocities could in principle be pattern motions which differ from the
actual bulk flow speed. But if the bulk flow were substantially more
relativistic than the observed motions, this would likely lead to
greater asymmetry between the approaching and the receding jet than is
observed in flux density (unabsorbed, as seen only at 43\,GHz, see
Sect.~\ref{sec:iongas}), arm-length, and speed.  If the velocities are
intrinsically the same in both jets, then the jets must be oriented
fairly close to the plane of the sky; a lower limit to the jet
inclination of 57\deg\ is obtained by taking the opposite extremes
allowed by the $1\sigma$ errors for the relativistic velocities in the
two jets: 0.090\,pc\,yr$^{-1}$ in the approaching jet, and
0.066\,pc\,yr$^{-1}$ in the receding jet.

\section{Ionised gas \label{sec:iongas}}

\begin{figure}[htbp]
    \begin{center}
    \includegraphics[width=6.5cm,angle=-90,clip]{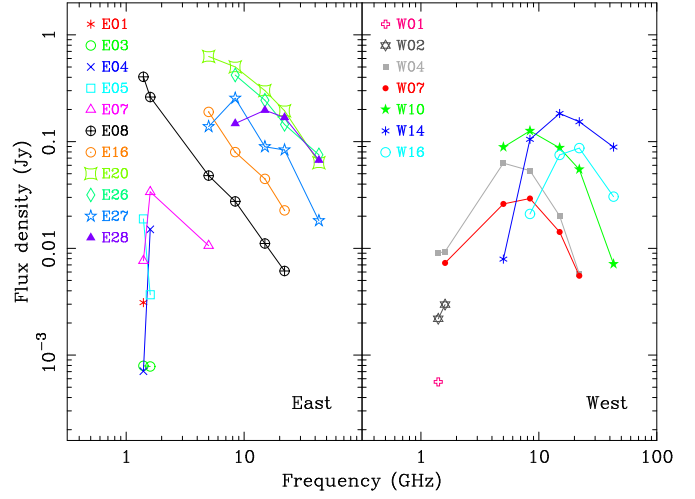}
    \caption{Spectra for selected components at epoch 1997.52.}
\label{fig:ffcomp}
\end{center}
\end{figure}

\begin{figure}[htbp]
    \begin{center}
    \includegraphics[width=6.5cm,angle=-90,clip]{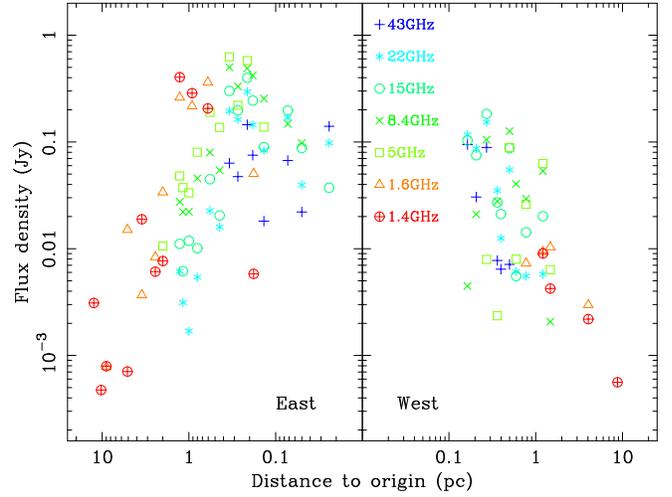}
    \caption{
        Flux densities for all of the components at epoch 1997.52
        presented in Fig.~\ref{ffmods}, plotted at their appropriate
        distances, and colour-coded by frequency to show the changing
        spectral shapes.
            }
      \label{fig:ffflux}
    \end{center}
\end{figure}

\begin{figure}[bth]
    \begin{center}
    \includegraphics[width=6.5cm,angle=-90,clip]{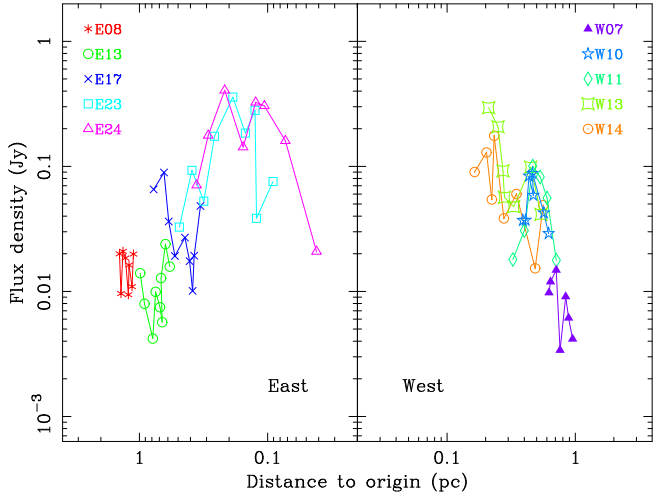}
    \caption{
        Flux density changes with time for some Gaussian components
        fitted to the 15\,GHz datasets, shown as a function of their
        distance to the origin to illustrate patchiness of the
        absorbing screen.
            }
\label{fig:RS}
\end{center}
\end{figure}

Figures~\ref{fig:ffims} and \ref{ffmods} show VLBA images and models at
43, 22, 15, 8, 5, 1.6, and 1.4\,GHz, all observed within a few days of
each other at epoch 1997.52 (Table~\ref{table:obstab}). Individual
component spectra are shown in Fig.~\ref{fig:ffcomp}; it is obvious
that there is a very wide range of spectral shapes, which proceed from
steep, through convex, to highly inverted, from the outer jets towards
the middle. Figure~\ref{fig:ffflux} provides a good illustration of
this general dependence of spectral shape on distance from the centre
(a well-defined location thanks to the tie-in to the kinematic analysis
through the co-eval 15\,GHz dataset).

On both sides of the gap there are components with a low-frequency
surface brightness well below 10$^{10}$\,K, and a low-frequency
spectral cutoff steeper than $\alpha=3$ if expressed as a power law.
As we have first reported in K99, we believe that the only
plausible explanation for these sudden truncations is free-free
absorption from ionised gas somewhere along the line of sight to the
radio jets, which, in its simplest form, leads to an exponential
spectrum; Ka01 came to the same conclusion based on three
frequencies observed at a later epoch. On the other hand, other
components have more gently convex spectra, and brightness temperatures
in excess of 10$^{10}$\,K, quite compatible with typical synchrotron
self-absorbed inner jet components of AGN\null. It is likely that
(intrinsic) synchrotron self-absorption and (external) free-free
absorption both play a role in fixing the observed spectral shape of
individual jet components of \object{NGC\,1052}, and if conditions are
such that similar peak frequencies result from both effects, it is
difficult to disentangle the two.

As a result of the free-free absorption, the images of
Fig.~\ref{fig:ffims} show, in going from high to low frequency, the
opening up of a distinctive hole in the middle of the radio
structure. Note that in contrast to Ka01, with our unambiguous
localisation of the kinematic centre we can ascertain that none of the
components visible at any frequency lower than 43\,GHz corresponds to
the core. At 43\,GHz the source structure appears to be fairly
symmetric, which, like the similarity between the velocities on both
sides, suggests that relativistic beaming effects are small, and that
the jets are therefore probably directed fairly close to the plane of
the sky. While at lower frequencies the hole is deepest over the
central region, the attenuation is much more pronounced along the
western jet than at corresponding locations along the eastern jet. At
$\sim$0.5\,pc along the eastern jet, and 1--2\,pc along the western
jet, the signature of free-free absorption becomes more and more
difficult to distinguish from plausible intrinsic,
synchrotron-self-absorbed, spectral shapes.

As we have previously reported (K99), the most straightforward way
to explain the observations is if the eastern jet is approaching us,
and the western jet receding from us, so that the latter gets covered
more deeply behind the free-free absorbing medium, almost regardless of
the detailed geometry of that medium.  Substantial coverage of the
approaching jet is obtained naturally if the absorber is geometrically
thick. It could very well be a thick disk- or torus-like region, as
postulated in many AGN unification models, oriented more or less
perpendicular to jets directed fairly close to the plane of the sky.

Variations on this scenario with a thinner disk or torus, either not
oriented orthogonal to the jets, or warped, are possible. Furthermore,
there is quite a variety of spectral shapes in our 1997.52 dataset
(Fig.~\ref{fig:ffcomp}), and the observed flux density variations of
individual components tracked over time at 15\,GHz (illustrated in
Fig.~\ref{fig:RS}) shows that the absorbing region, while possessing a
fairly well-defined overall geometry, has substantial patchiness in
detail. We therefore think it is not very meaningful to use a single
epoch to enter into detailed modeling of the (a)symmetry between the
jet flux densities, in order to try to derive more stringent bounds on
the jet and screen geometry, or to attempt spectral decompositions into
intrinsic shapes and absorption contributions, in an effort to find
detailed radial or transverse opacity profiles. We plan to improve this
situation with multi-epoch multi-frequency monitoring, in which we hope
to use a series of moving components to trace out overlapping parts of
the profile.

\begin{figure}
    \begin{center}
    \includegraphics[width=6.8cm,angle=-90]{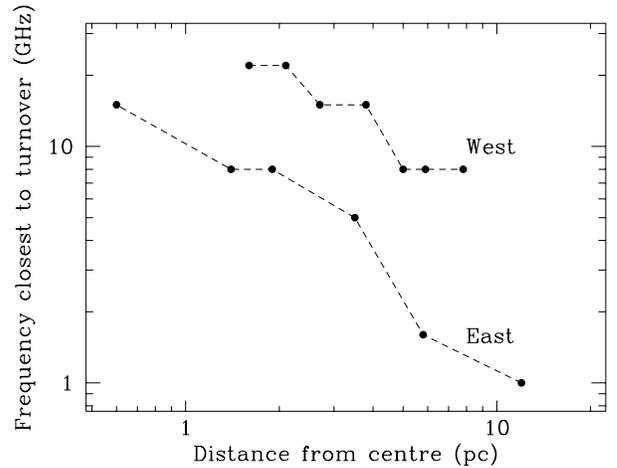}
    \caption{
        Peak frequency, approximated as the frequency at which
        the component is brightest, as a function of component distance
        from the centre.
            }
\label{fig:ffrad}
\end{center}
\end{figure}

In order to gain an impression of the overall trend of free-free
absorption opacity as a function of distance from the centre, we show
in Fig.~\ref{fig:ffrad} the frequency at which components with a convex
spectrum have their greatest flux density; this should be close to the
turnover frequency which would be found if a detailed spectral
decomposition were possible. The turnover frequency decreases roughly
with distance along the jets to the power 0.7--1, and is fairly
consistently a factor of 4--5 lower along the eastern jet than at the
corresponding distance along the western jet.  We compute from
$\tau\sim1$ at 43\,GHz (the highest turnover frequency seen) that the
volume density would be $n_{\rm e}\sim10^5$\,cm$^{-3}$ if the free-free
absorbing gas would be distributed uniformly along a path-length of
0.5\,pc with a temperature $T=10^4$\,K.

\section{Atomic gas \label{sec:atogas}}

\begin{figure} \begin{center}
    \includegraphics[width=7cm,angle=-90,clip]{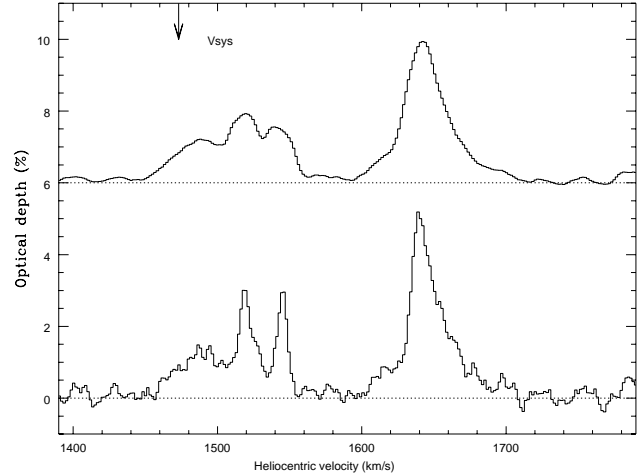}
    \caption{
        Integrated H{\sc i} optical depth profile observed in
        1998.  Bottom: at the full spectral resolution of
        2\,km\,s$^{-1}$. Offset: smoothed to 20\,km\,s$^{-1}$ for
        comparison to the spectrum of vG86.
            }
\label{hioptd}
\end{center}
\end{figure}

\begin{figure} 
\begin{center}
    \includegraphics[width=8cm,angle=0,clip]{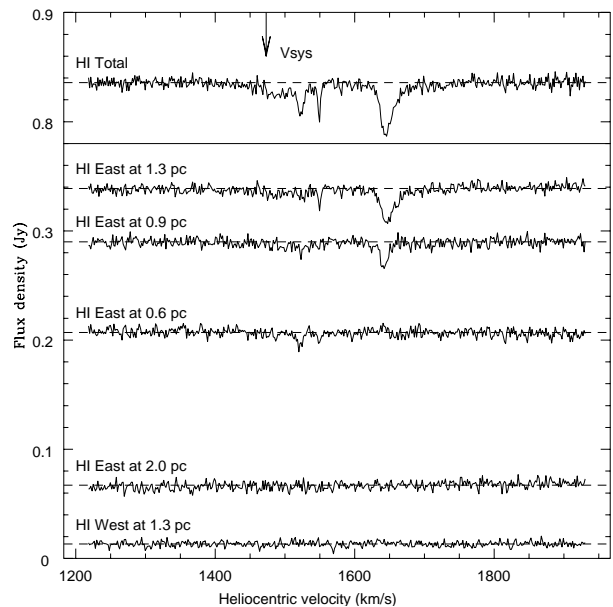}
    \caption{
        H{\sc i} absorption spectra observed in July 1997 at various
        locations along the jets of \object{NGC\,1052}. Note the offset
        flux density scale for the summed spectrum.
            }
\label{hispec97}
\end{center}
\end{figure}

\begin{figure}[htbp] 
\begin{center}
    \includegraphics[width=8cm,clip]{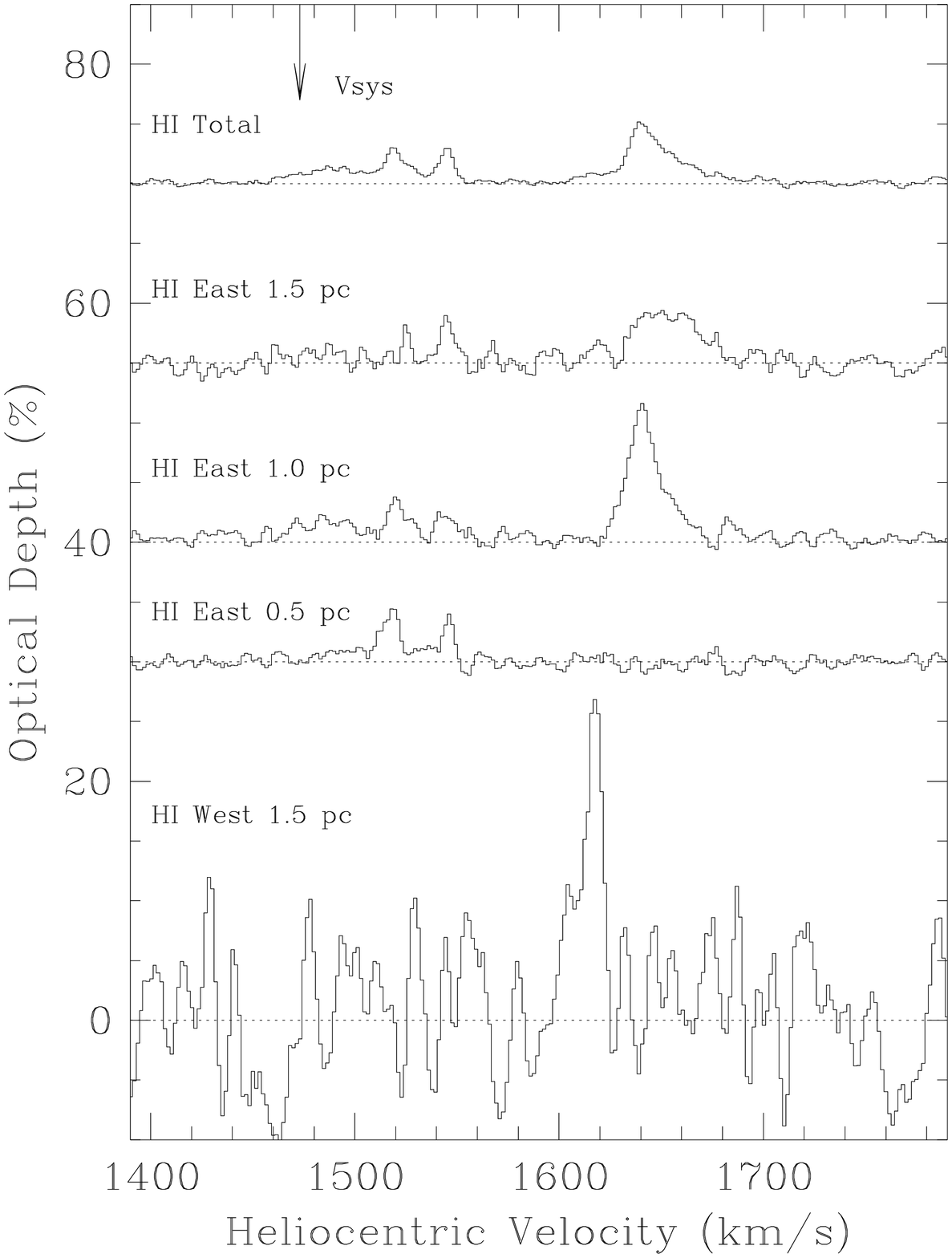}
    \caption{
        H{\sc i} optical depths observed in 1998 at various locations
        along the jets of \object{NGC\,1052}. Multiple offsets are
        used for clarity, with dotted lines to show the zero levels.
            }
\label{hispec98}
\end{center}
\end{figure}

Figure~\ref{hioptd} shows that the integrated H{\sc i} absorption
starts around 1440\,km\,s$^{-1}$, just blueward of the systemic
velocity, $v=1474$\,km\,s$^{-1}$, and then continues towards higher
velocities (redshifts) over more than
300\,km\,s$^{-1}$. Figures~\ref{hispec97} and \ref{hispec98} depict
spatially resolved H{\sc i} spectra from July 1997 and 1998, in flux
density and optical depth, respectively, at all locations where the
signal-to-noise ratio allowed detection of line absorption at the level
of a few percent. The spectrum of the western jet is shown for both
epochs, but its flux density is so severely attenuated by free-free
absorption that it was not possible to detect H{\sc i} absorption with
the modest sensitivity of the 1997 VLBA-only data. The more sensitive
1998 dataset does reveal very high opacity (up to 25\%) absorption
towards the brightest part of the western jet, situated about 1.5\,pc
from the centre. We did not detect any line features at 2\,pc or larger
distances along the eastern jet; Fig.~\ref{hispec97} shows, as an
illustration, the 1997 spectrum at 2\,pc, where the 2$\sigma$ optical
depth limit is 2\%\ for features with 20\,km\,s$^{-1}$ FWHM. We believe
that there are likely to be three absorption systems with different
characteristics, at least two of which are probably due to atomic gas
on parsec or sub-parsec scales, local to the AGN environment, rather
than distributed on galactic scales.

The most distinct, deepest absorption occurs above 1600\,km\,s$^{-1}$.
From the 1998 dataset, we see in Fig.~\ref{hispec98} that the western
jet shows a 20--25\%\ deep line centred at 1620\,km\,s$^{-1}$, with a
FWHM of about 20\,km\,s$^{-1}$ and a possible blue wing. At 1\,pc along
the eastern jet there is a roughly 10\%\ deep peak near
1640\,km\,s$^{-1}$ with a FWHM of some 18\,km\,s$^{-1}$, while at
1.5\,pc east of the centre there is a 5\%\ deep feature centred near
1652\,km\,s$^{-1}$ with a FWHM of 30\,km\,s$^{-1}$. The 1997 dataset
(Fig.~\ref{hispec97}) gives similar results for the eastern jet
although the details are dependent on the precise locations of the
background features. We have the impression that these features are all
part of a single "high velocity system". This absorber then has a
west-to-east velocity gradient of some 10\,km\,s$^{-1}$\,pc$^{-1}$,
with the possible blue wing in the west and the increasing velocity
width in the east perhaps indicating an even larger velocity gradient
towards its outer edges; these could be the tell-tale signs of a
rotating structure. The high velocity system is absent at 2\,pc and
larger distances, as far as can be determined along the eastern jet.
Even more interestingly, while the high velocity system has prominent
H{\sc i} absorption along both the receding and the approaching jet, it
appears to have a central hole: there is no high velocity absorption
deeper than 0.4\%\ (2$\sigma$ limit) towards the relatively bright
innermost jet component, centred 0.6\,pc east of the core. This shows a
lack of neutral gas in the central parsec around the AGN, while, as
discussed in Sect.~\ref{sec:discussion}, both ionised and molecular gas
do exist close to the centre.

By contrast, there is a ``low velocity system", which is fairly smooth
but relatively shallow, with optical depth 1--2\%. It spans
1440--1570\,km\,s$^{-1}$, asymmetrically straddling the systemic
velocity $v_{\rm sys}=1474$\,km\,s$^{-1}$, determined from galactic
absorption lines (\saruu). Since it is found equally in all jet
components where the signal-to-noise ratio is adequate, the low
velocity system could either arise near the AGN or in gas on much
larger than parsec scales.

Superimposed, in the velocity range 1500--1570\,km\,s$^{-1}$ are some
sharper features, having widths of 3--15\,km\,s$^{-1}$ and depths up to
5\%, which predominate at quite specific locations along the inner
2\,pc of the eastern jet. The most prominent one, centred at
1522\,km\,s$^{-1}$ with a FWHM of 6\,km\,s$^{-1}$, is seen mostly at
0.6\,pc from the core, where it raises the local total absorption depth
to 10\%.  Remarkably, a sharp absorption line near
1549--1550\,km\,s$^{-1}$ is evident both 0.6\,pc and 1.3\,pc east of
the centre, giving an optical depth of 5--6\%, but there is no line
deeper than 2\%\ (2$\sigma$ limit) near this velocity towards the
intervening eastern jet component at 0.9\,pc. It is also noteworthy
that these sharpest features (as ascertained from our higher resolution
1998 dataset) are remarkably kinematically quiescent, with a FWHM of
only 3--5\,km\,s$^{-1}$.

The full velocity extent of the absorption is roughly similar to that
found in the 1982/3 VLA spectrum obtained by vG86. However, the
detailed profile in the lower velocity range has evolved markedly over
time, as we illustrate by convolving our 1998 integrated spectrum to
the ten times lower spectral resolution of the older data
(Fig.~\ref{hioptd}). Furthermore, the peak absorbed flux density, which
occurs in the high velocity system at both epochs, has increased to
about 45\,mJy in our integrated 1997 profile, as compared to about
15\,mJy in 1982/3, while the total flux densities are quite comparable:
840\,mJy in 1982/3, compared to 950\,mJy in 1997 summed over all VLBI
components (the recent L-band WSRT observations show that the
arcsecond-scale lobes in addition contain less than 50\,mJy
each). Thus, the peak optical depth, computed from the integrated flux
densities, was 5\%\ in 1997, but only 2\%\ in 1982/3. \shoiep\ have
published earlier data, from 1979 using the old WSRT line backend, and
from 1981/2 with the VLA. They report a fairly consistent continuum
flux density of 800\,mJy in 1979, and a peak optical depth of only 1\%,
but we have the impression that, in view of the lower sensitivity and
possible spectral baseline uncertainties evident in their figures, the
peak depth is not inconsistent with the 2\%\ seen in 1982/3 by vG86.
However, even despite the limited sensitivity and spectral resolution,
it looks like in the oldest spectra, the absorption in the lower
velocity range might be at least as deep as the high velocity
absorption, whereas in both the 1982/3 spectrum and in our 1998 data
convolved to 20\,km\,s$^{-1}$ resolution, the high velocity system
optical depth is at least a factor of 2 in excess of the optical depth
at lower velocities.  All of these changes can be easily understood
since the absorption at the different velocities covers regions of the
jet which have a typical size of 1\,pc or even rather less, and that
the emitting components move with a velocity of $0.26c$
(Sect.~\ref{sec:kinematics}), so that the continuum in the background
of an absorber is ``refreshed'' entirely in a decade or less.

Optical depths ranging from 2\%\ to 20\%, and a typical FWHM of
20\,km\,s$^{-1}$, imply column depths of $N_{\rm H}=10^{20}$--$10^{21}
T_{\rm sp,100}$\,cm$^{-2}$, and hence densities of $n_{\rm H} =
100$--$1000\,T_{\rm sp,100}$\,cm$^{-3}$ if there were a uniformly
filled path-length of 0.5\,pc, comparable to the transverse extent of
many individual absorbers. The normalisation here adopts a typical
Galactic spin temperature, $T_{\rm sp}=100$\,K, but conditions close to
an AGN (e.g., \maloy) may well raise this by one or two orders of
magnitude.

\section {Molecular gas \label{sec:molgas}}

\begin{figure}[htbp]
\begin{center}
    \includegraphics[width=6.8cm,angle=-90,clip]{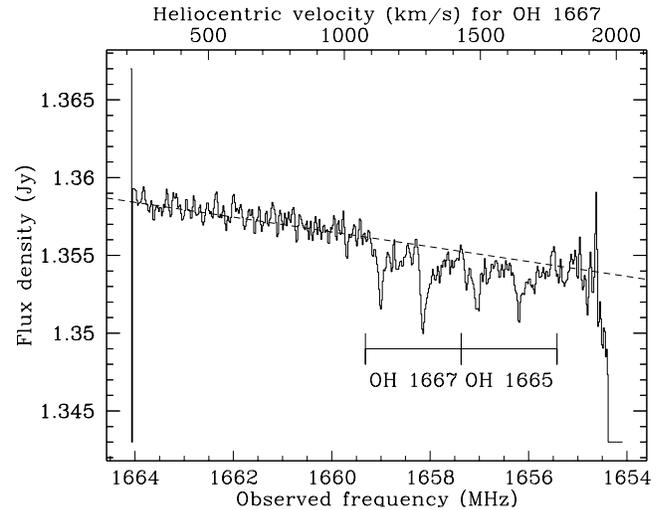}
    \caption{
        A single WSRT spectrum with the velocity range
       1430--1780\,km\,s$^{-1}$ indicated for both the 1667 and the
       1665\,MHz OH main line.
            }
\label{ohmain}
\end{center}
\end{figure}

\begin{figure}[htbp] 
\begin{center}
    \includegraphics[width=8cm,clip]{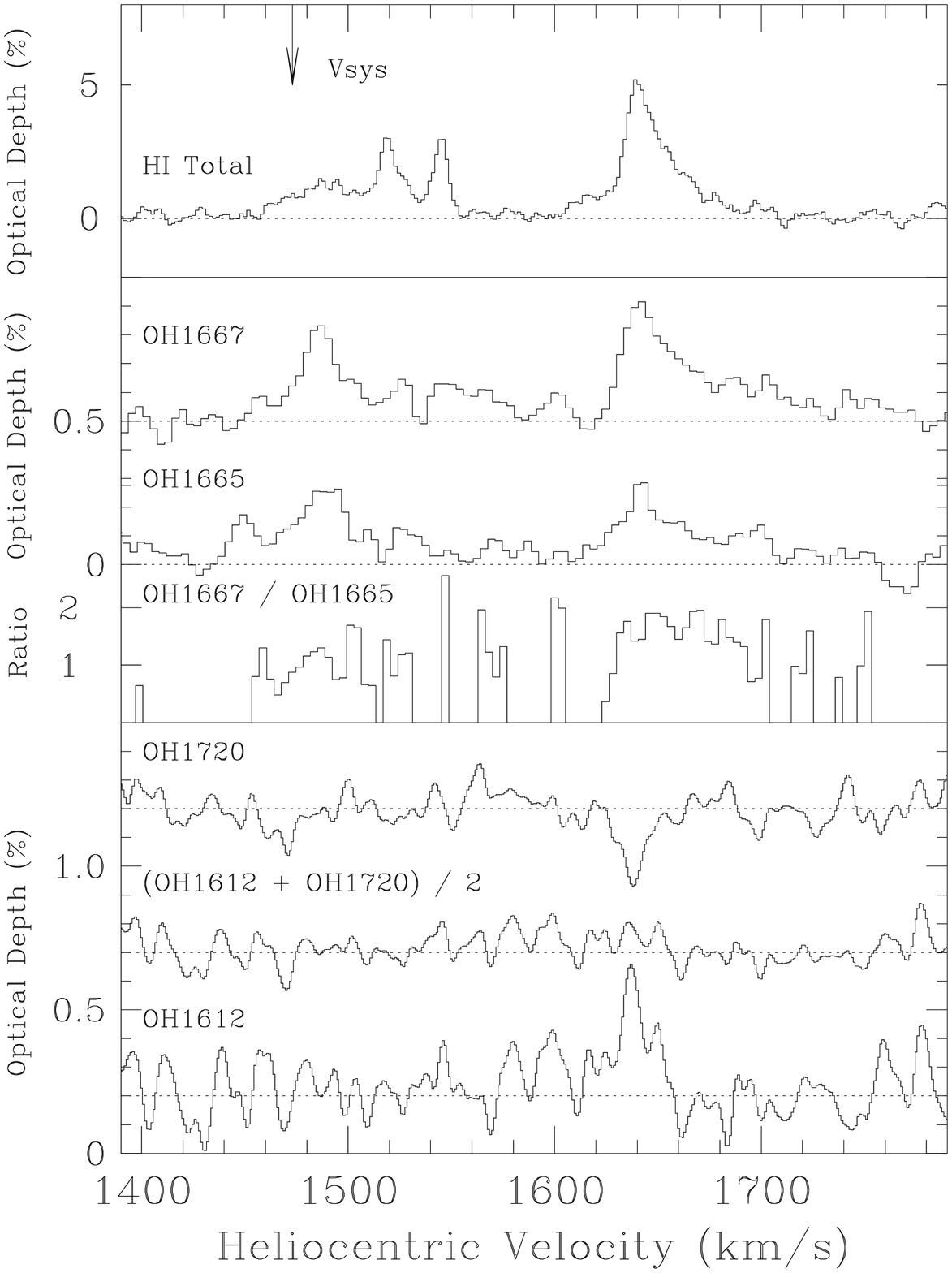}
    \caption{
        WSRT optical depth measurements (negative is emission) for all
        four 18\,cm OH lines. The offset zero lines are indicated; also
        note the different scale for our integrated VLBI H{\sc i}
        spectrum, shown convolved to the same resolution for
        comparison.
            }
\label{ohall}
\end{center}
\end{figure}

It is clear from Figs.~\ref{ohmain} and \ref{ohall} that while the
deepest OH absorption feature, $\sim$0.4\% in the 1667\,MHz line near
1640\,km\,s$^{-1}$, corresponds to that originally detected by \omapwp,
18\,cm OH is in fact detectable over a wide range in velocity. The
frequency difference of 2.0\,MHz between the 1667 and 1665\,MHz main
lines could potentially be confused with a velocity difference of
350\,km\,s$^{-1}$ in the same line. That is somewhat more than the
velocity range seen in H{\sc i} absorption, and indeed, it seems from
Fig.~\ref{ohmain}, where the velocity interval 1430--1780\,km\,s$^{-1}$
is indicated for both lines, that it is fortunately just possible to
discriminate uniquely between the two lines. Having extracted both of
the main lines from the single WSRT spectrum shown in
Fig.~\ref{ohmain}, we see in Fig.~\ref{ohall} that their profiles match
strikingly well, which is further evidence that the velocity range over
which detectable OH occurs does not exceed 350\,km\,s$^{-1}$ and that
there is thus no significant overlap in the spectrum between the OH
main lines. The peak absorption depth in the 1667\,MHz line is about
0.4\%\ at 1640\,km\,s$^{-1}$ and suggests a column depth of the order
of $10^{14}$\,cm$^{-2}$ if we take a line width of 20\,km\,s$^{-1}$ and
$T_{\rm ex}=10$\,K.  Figure~\ref{ohall} shows that the ratio between
the OH 1667 and 1665\,MHz lines ranges from near 1 at low velocities to
approximately 2 in the high velocity system. The ratio in local thermal
equilibrium (LTE) would be 1.8, but at the densities that prevail in
the interstellar medium it is expected that OH is hardly ever
thermalised.

We have also been able to detect with high confidence the satellite
18\,cm OH lines, 1612 and 1720\,MHz, in the high velocity system,
around 1640--1650\,km\,s$^{-1}$; the profiles are shown in
Fig.~\ref{ohall} along with the main lines. While, just like the main
lines, the 1612\,MHz line is in absorption, with a peak optical depth
of about 0.4\%, the 1720\,MHz line is in emission, with about equal
strength !  There is some evidence in the spectrum for the occurrence
of the 1720\,MHz line in the low velocity system, around
1470\,km\,s$^{-1}$, again in emission, but we regard this as quite
tentative. Our sensitivity to features in the 1612\,MHz line at these
velocities was decreased by RFI{}. Thus, the two satellite OH lines
probably have conjugate (i.e.\ mirror image emission and absorption)
profiles. This has been known in Galactic conditions and has been
observed in \object{Cen~A} (\vanot) and \object{NGC\,253}
(\fraoi). Generally, it results from excitation of OH molecules in a
far infra-red (FIR) radiation field when the OH column density is
sufficiently large and the FIR OH lines become optically thick. The
behaviour was successfully modelled for \object{Cen~A} in a non-LTE
approach by \vanotp; competing pumping mechanisms determine which line
is in emission and which one in absorption in specific density and
temperature regimes. It appears that the \object{NGC\,1052}
observations, where the main lines have a ratio close to the LTE value
and the satellite lines show conjugate behaviour, resemble the
conditions close to the \object{Cen~A} circumnuclear disk. In
\object{NGC\,1052} there is also a circumnuclear infrared excess
(Becklin et al.\ \cite{bec82}), and we favour an interpretation in
which this OH absorption occurs in the nuclear region. We plan to model
this in more detail when VLBI spectral imaging of the OH lines is
available to constrain the geometry.

\section{Discussion \label{sec:discussion}}

The general picture which emerges from the data presented in the previous
sections supports the scenario often invoked in AGN (unification)
models. The central engine produces twin radio jets, which lie along the polar
axis of a toroidal or disk-like region into which matter is
accumulating. However, some important questions remain.

The jet velocity in NGC\,1052 is only mildly relativistic, rather than
approaching the velocity of light, as in many other AGN.  Our
observations may refer only to a slowly moving external surface of the
jets, so that the radiation from a faster moving central spine is
undetectable because it is beamed away from the line of sight. However,
NGC\,1052 is one of the lowest luminosity galaxies in which a jet
velocity has been measured to date. It is quite possible that low
luminosity objects intrinsically have slower jets on average.

The combination of free-free absorption, from a geometrically thick,
albeit patchy, ionised structure covering the inner parsec, and H{\sc
i} atomic gas predominantly in an annulus 1--2\,pc around the centre,
is quite natural in principle. The innermost region and/or the surface
of the accretion disk or torus receive the most intense radiation from
the AGN, and thus have the largest degree of ionisation. Indeed, recent
work by \weaoop, \guaoop, \guappp, and Kadler et al.\ (\cite{kad02a})
has shown the presence of large column depths ($N_{\rm
H}$=10$^{22}$--10$^{24}$\,cm$^{-2}$) of soft X-ray absorbing gas along
our line of sight to the nuclear continuum source. Furthermore, 
some models of the X-ray spectral details suggest that the distribution
of the absorbing gas is patchy, or that two components with a
substantially different density are involved, matching the evidence we
see for patchiness also in the radio spectra and flux density evolution
of different jet components at various distances from the core. The
detection by \beaoop\ of a broad H$\alpha$ emission line, visible only
in polarised (scattered) light, is also evidence for obscuration by a
canonical equatorial torus or disk. But several inconsistent reports on
the (non)detection of CO emission and absorption (\weaot; \wikot;
\knaoy) have left the amount of molecular gas near the nucleus
uncertain.

However, the nature of the high velocity H{\sc i} absorption system is not
clear. The apparent 10\,km\,s$^{-1}$\,pc$^{-1}$ velocity gradient could
correspond to a rotation component of the motion. Naively, for matter at a
distance of 1\,pc from the core, this rotation would imply a modest enclosed
mass of about 10$^7$\,M$_{\odot}$, but we suspect that the actual geometry is
rather more complex. In particular, this system is redshifted by
130--200\,km\,s$^{-1}$ compared to the stellar recession
velocity of NGC\,1052. Since the
spatial distribution strongly suggests that the gas is quite local to the AGN
environment, it is most likely fairly cohesively falling in to the
nucleus. One might speculate that this is the result of recent accretion of
gas due to (a series of) mergers.  NGC\,1052 has complex, probably triaxial
kinematics on arcsecond scales, as revealed by optical spectroscopy (\plaoi)
and radio observations of H{\sc i} (vG86), and it is not obvious how these
large-scale motions connect to those of parsec-scale features. 
Under an inflow
scenario, we are puzzled by the absence of H{\sc i} absorption towards 
the innermost
approaching jet, given that the three dimensional distance to the ionising
central source may then not differ very much from that of gas seen 1--2\,pc
away from the nucleus in projection on either side. 
We point out, incidentally,
that the radio jets do not appear to be ``significantly'' oriented with
respect to the isophotal or rotation axes of either the stellar body or the
neutral or ionised gas in NGC\,1052, despite early suggestions to the contrary
(\jeair; \daviy; vG86). While the initial direction of ejection is most
likely determined by the rotation of the black hole, it is unclear what
determines the subsequent bending.

While the WSRT observations of the 18\,cm OH lines in NGC\,1052 do not
resolve the arcsecond-scale source, we believe that it is likely that
in the high velocity system the OH lines are due to molecules situated
towards the eastern, approaching jet, at distances of 1--2\,pc from the
core, just like the H{\sc i} line. We know from our 1997 18\,cm
continuum VLBA data that the receding jet contained a flux density of
only $\sim$20\,mJy, so that the the local OH opacity in the 1667\,MHz
line would have to be as high as 25\%\ if the absorption were to occur
only on that side. Just as importantly, the high velocity part of the
OH profile matches that of the integrated H{\sc i} absorption rather
well (Fig.~\ref{ohall}). The peak H{\sc i} absorption depth is 5\% of
the total flux density, as compared to 0.4\%\ in OH. Remembering that
this is based on values integrated over the source, and with a 3.5 year
time difference, we find that the opacity ratio of 12:1, which converts
to a column depth ratio of $10^{6}$ (naively taking $T_{\rm sp}=100$\,K
for H{\sc i} and $T_{\rm ex}=10$\,K for OH), matches very well the
models of \meaoyp\ for conditions near an AGN.

While we cannot rule out that the ``low velocity'' H{\sc i} and OH
absorption is the result of a long path length through the galaxy, the
OH excitation points toward an association of this gas with the nuclear
region.  The sharper, localised H{\sc i} absorption lines also have no
clear counterpart in OH, although the spectral resolution in OH was a
factor 2--4 lower. Perhaps these are clouds with a different ionisation
parameter and a lower fraction of molecular gas.

It is also possible that there are at least a few differently distributed
molecular gas components present in the circumnuclear medium. \claoip\ have
found H$_2$O maser emission, indicating the presence of molecular gas at
0.1--0.2\,pc to the west of the core, along the receding jet (note that the
zero point in the images of \claoip\ is not on the core, since they did not
have the benefit of multi-epoch and multi-frequency datasets). Possibly, these
masers are excited by interaction with the jets. We have no explanation,
however, for the fact that while the H$_2$O masers have the same speeds as the
``high velocity system'', they appear to have a different location than the
majority of the H{\sc i} and OH line producing gas.

Remarkably, therefore, in the complex accretion region, which the inner
parsec around the AGN surely is, it appears possible for some molecular
and atomic clouds to preserve a relatively quiescent existence, with
apparently highly ordered motions of the high velocity system, and
kinematically fairly undisturbed separate sharp H{\sc i} absorbing
clouds. There are other active galaxies in which somewhat similar
obscuring regions have been found, but they each seem to have their own
peculiarities. For example, \kaspwp\ find at least three components
with different ionisation parameters in X-ray absorption towards
NGC\,5548, with variable broad lines showing a location between
10$^{14}$ and 10$^{16}$\,cm. In NGC\,1275 VLBI data show that the
approaching jet is apparently not covered by free-free absorption, but
the inner portion of the counter-jet is hidden behind a thick or warped
disk (e.g. \walpp; \levot). In NGC\,4261 \jonpqp\ find evidence for a
geometrically thin obscuring disk. There is also H{\sc i} absorption in
NGC\,4261, but in this case, \vanppp\ have found that it occurs only
towards the counter-jet, and that the column depth is only $N_{\rm
H}=2.5\times10^{19}$\,cm$^{-2}$. As a final example, in NGC\,5793,
\pihppp\ find that there are H{\sc i} absorption components at several
velocities, which each cover a different VLBI continuum component, and
they span a region of 30\,pc, rather larger than in NGC\,1052. It is
also intriguing to extend the comparison with NGC\,5793 to molecular
gas. The 1667 and 1665\,MHz OH main lines have been imaged in
absorption by \hagppp.  Their extent is somewhat uncertain, but is at
least 5\,pc. Interestingly, there is a 9\,km\,s$^{-1}$\,pc$^{-1}$
velocity gradient across the source, but in this case it is centred on
$v_{\rm sys}$. There is also H$_2$O maser emission in NGC\,5793, but at
another position and a blue-shifted velocity (\diapq).

\section{Summary \label{sec:summary}}

We have presented and discussed multiple VLBA continuum and spectral
line imaging and WSRT spectroscopic observations of the compact
variable nuclear radio source in the nearby active galaxy NGC\,1052. Our
main conclusions can be summarised as follows:

1. The radio source consists of slightly curved bi-symmetric
jets with multiple sub-parsec scale features.

2. There are outward motions, reasonably linear on the sky and
constant over time, of typically $v_{\rm app}\sim0.26c$
(${\rm H}_0=65$\,km\,s$^{-1}$\,Mpc$^{-1}$) on each side.

3. Symmetry shows that the jets are oriented near the plane of the sky.

4. Spectral shapes of jet components ranging from steep, through
convex, to highly inverted suggest that synchrotron self-absorption and
free-free absorption are both at play along the inner parsecs of the jets.

6. Free-free absorption leads to an asymmetric central gap opening up
towards lower frequencies. Thus, the western jet is receding, and is
partially obscured along at least 1\,pc.

7. The eastern jet is approaching, but is still partially covered along
approximately 0.3\,pc. This probably indicates a geometrically thick
absorber, oriented roughly orthogonal to the jets.

8. Globally, the free-free opacity drops with increasing distance from
the centre, but component variability suggests patchiness of the
absorber. The highest turnover frequency of 43\,GHz indicates a volume
density of $n_{\rm e}\sim10^5$\,cm$^{-3}$ if the free-free absorbing
gas were distributed uniformly along a path-length of 0.5\,pc.

9. H{\sc i} atomic gas absorption is distributed in front of the
approaching as well as the receding jet and has sub-pc scale
structure. We distinguish three absorption systems with different
characteristics.

10. ``High velocity'' H{\sc i} gas, receding by 125--200\,km\,s$^{-1}$ with
respect to the systemic velocity of NGC\,1052
($v_{\rm sys}=1474$\,km\,s$^{-1}$), shows the most prominent absorption.

11. The high velocity H{\sc i} absorber may have a continuous velocity
gradient across the nucleus of some 10\,km\,s$^{-1}$\,pc$^{-1}$, to
connect a 20--25\%\ deep feature around 1620\,km\,s$^{-1}$ 1.5\,pc away
from the core along the receding jet, with 5--10\%\ deep features near
1640 and 1650\,km\,s$^{-1}$ located 1 and 1.5\,pc away from the core
along the approaching jet.

12. The innermost parsec (of the approaching jet) shows a deficit of
this high-velocity component, which shows that it is local to the
AGN. The atomic gas in the ``central hole" may be largely ionised due
to the proximity to the AGN; this is the location of the deepest
free-free absorption.

13. A broad, shallow H{\sc i} absorption component, with a peak depth
of about 2\%, spans the ``low velocity'' range
1440--1570\,km\,s$^{-1}$, asymmetrically straddling the systemic
velocity. It covers all jet components in which we have the sensitivity
to detect it, and could arise either in gas local to the AGN or on
galactic scales.

14. Sharper H{\sc i} features, having widths of 3--15\,km\,s$^{-1}$ and
depths up to 5\%, are superimposed in the range
1500--1570\,km\,s$^{-1}$. These features each extend over only a few
tenths of a pc and can be detected at various places along the inner
2\,pc of the approaching jet.

15. Since the absorbing neutral gas has structure on scales of 0.1--1\,pc,
back-lit by compact, near-relativistically moving features in the
jets, it is logical that the overall H{\sc i} absorption line profile
of our data differs significantly from that in VLA observations taken
15 years earlier.

16. The 1667 and 1665\,MHz 18\,cm OH main lines are clearly present in
absorption along the full velocity span seen also in H{\sc i}, roughly
1440--1670\,km\,s$^{-1}$. The peak absorption depth in the 1667\,MHz
line is about 0.4\%\ at 1640\,km\,s$^{-1}$, suggesting a column depth
of order $10^{14}$\,cm$^{-2}$. The ratio between the OH 1667 and
1665\,MHz lines ranges from near 1 at low velocities to approximately 2
in the high velocity system.

17. The 1612 and 1720\,MHz 18\,cm OH satellite lines have also been
detected, in the high velocity system at 1630--1650\,km\,s$^{-1}$. The
1612\,MHz line is in absorption, but the 1720\,MHz line is in emission,
both with a peak strength near 0.25\%. Their conjugate profiles
resemble the behaviour in \object{Cen~A} and \object{NGC\,253}.

18. The OH main line and the total H{\sc i} profiles are remarkably
similar in the high velocity system, and we suggest co-location of
these high velocity atomic and molecular gas components. They probably
do not coincide with the H$_2$O masers at 0.1--0.2\,pc along the
receding jet, even though these are at the same velocity.

19. Other questions also remain regarding the nature of the high
velocity system. Interpretation of the velocity gradient in H{\sc i} as
evidence for a structure rotating around the nucleus is contradicted by
the fact that the centroid is redshifted by 150\,km\,s$^{-1}$ or more
from the systemic velocity. But if it is instead infalling gas, the
nature of the central hole in H{\sc i} is unclear.

\begin{acknowledgements}
  We thank the staff of the VLBA and the WSRT for their extensive
  support.  The VLBA is an instrument of the NRAO, a facility of the
  National Science Foundation of the USA, operated under cooperative
  agreement by Associated Universities, Inc.  The WSRT is operated by
  ASTRON (The Netherlands Foundation for Research in Astronomy) with
  support from the Netherlands Foundation for Scientific Research
  (NWO).  This work is also based on observations with the 100-m
  telescope of the MPIfR (Max-Planck-Institut f\"ur Radioastronomie) at
  Effelsberg.  This work was initially supported in part by the US
  National Science Foundation under grant AST-9420018, and has made use
  of NASA's Astrophysics Data System (ADS), and the NASA/IPAC
  Extragalactic Database (NED) which is operated by the Jet Propulsion
  Laboratory, California Institute of Technology, under contract with
  NASA. We are grateful to Frodo Wesseling for assistance with the data
  reduction, and to Dan Homan, Matthias Kadler and Matt Lister for
  helpful discussions and suggestions. We thank Gabriele
  Giovannini for prompt and helpful comments as referee.
\end{acknowledgements}

\end{document}